# An Introduction to Implicit Regression: Extending Standard Regression to Rotational Analysis and Non-Response Analysis

R. D. Wooten, Ph.D.

Department of Mathematics and Statistics

University of South Florida, Tampa

**Abstract**: Statisticians usually restrict regression to model relationships that are explicitly defined dependent and independent random variables; this paper outlines the newly developed method of **non-response analysis** and **rotational analysis** for evaluating co-dependent variables without an obvious subject response. The concepts outlined challenge the notion of fixed effects; unity is included as a random measure (variable) ignoring the assumption of independence and the **degree of separation** is outlined which is a measure of model quality.

## 1.0 Introduction to Implicit Regression

**Implicit Regression** was developed by R. D. Wooten to address co-dependent relationships among measured variables with normal random error. Consider the equation containing exactly two variables of interest,

$$g(x, y) = h(x, y|\theta), \tag{1}$$

where $g(x, y)$ is a fixed function with well-defined constant coefficients and $h(x, y)$ is defined in terms of the unknown coefficients, $\theta = \{\alpha_0, \alpha_1, \alpha_2, \ldots, \alpha_m\}$.

The observed data gives us a fixed value $z_i = g(x_i, y_i)$ and

$$g(x_i, y_i) = h(x_i, y_i|\theta) + \omega_i, \tag{2}$$

where $\omega_i$ is assumed to be normally distributed with mean of zero (0) and constant variance $\sigma^2$; that is,

$$\omega_i \sim N(0, \sigma^2).$$

However we make no assumption of the underlying distribution of $x \sim F_X(x)$ and $y \sim F_Y(y)$, and allow random error in both readings, $x_i = x + \delta_i, \delta \sim N(0, \sigma_\delta^2)$ and $y_i = y + \varepsilon_i, \varepsilon \sim N(0, \sigma_\varepsilon^2)$.

Using ordinary least squares, we can construct a system of equations that, depending on the tractability of the functions and their derivatives, can be solved to find parameter estimates of the unknown coefficients, $\theta = \{\alpha_0, \alpha_1, \alpha_2, \ldots, \alpha_m\}$.

Equations take the form:

$$\sum_{i=1}^{n}\left[g(x_i, y_i) \times \frac{\partial h}{\partial \alpha_j}\right] = \sum_{i=1}^{n}\left[h(x_i, y_i|\theta) \times \frac{\partial h}{\partial \alpha_j}\right], j = 0,1,2,\ldots m. \quad (3)$$

Assume that $h(x, y|\theta)$ is an additive model taking the form

$$h(x, y|\theta) = \alpha_0 + \alpha_1 T_1(x, y) + \cdots \alpha_m T_m(x, y),$$

where $T_i(x, y) = x^{a_i} y^{b_i}$, where the exponents are constants, $a_i, b_i \in R$; and $g(x, y)$ a function with fixed coefficient that does not contain terms found in $h(x, y)$.

This construction allows for three main types of analysis: **Rotational Analysis** (Wooten, 2013 & 2015), **Non-Response Analysis** (Wooten, 2011, 2015 & Allison, 2013), and **Standard Regression**.

### 1.1 Rotational Analysis

In **rotational analysis**, let the terms of interest be the set $\{T_1(x, y), \ldots, T_m(x, y)\}$; to test the relationship one term at a time by considering the implicit model such that

$$g(x, y) = T_j(x, y)$$

and

$$h(x, y|\theta) = \alpha_0 + \sum_{k \neq j} \alpha_k T_k(x, y).$$

Then the system of equations becomes

$$\sum_{i=1}^{n}[T_k(x_i,y_i) \times T_j(x_i,y_i)] = \sum_{i=1}^{n}\left[\left(\alpha_0 + \sum_{k \neq j}\alpha_k T_k(x,y)\right) \times T_k(x_i,y_i)\right]; k = 1, \ldots, m, k \neq j$$

and

$$\sum_{i=1}^{n}T_j(x_i,y_i) = n\alpha_0 + \sum_{i=1}^{n}\sum_{k \neq j}\alpha_k T_k(x_i,y_i),$$

which allows each term to be tested in relation to the other remaining terms.

For example, if the terms of interest are $\{x, y, xy\}$, then there are three rotations:

$$y = \alpha_0 + \alpha_1 x + \alpha_2 xy$$

$$x = \alpha_0 + \alpha_1 y + \alpha_2 xy$$

$$xy = \alpha_0 + \alpha_1 x + \alpha_2 y,$$

which can be used to analyze the nature of the relationship that exist between the underlying variables. For the first rotation, the system of equations takes the form:

$$\sum_{i=1}^{n}y_i = n\alpha_0 + \alpha_1 \sum_{i=1}^{n}x_i + \alpha_2 \sum_{i=1}^{n}x_i y_i$$

$$\sum_{i=1}^{n}x_i y_i = \alpha_0 \sum_{i=1}^{n}x_i + \alpha_1 \sum_{i=1}^{n}x_i^2 + \alpha_2 \sum_{i=1}^{n}x_i^2 y_i$$

$$\sum_{i=1}^{n}x_i y_i^2 = \alpha_0 \sum_{i=1}^{n}x_i y_i + \alpha_1 \sum_{i=1}^{n}x_i + \alpha_2 \sum_{i=1}^{n}x_i y_i$$

which can be scrutinized using standard regression methods in turn.

In each of these rotations, both variables are tractable; in the first rotation $y = \frac{a_0 + a_1 x}{1 - a_2 x}$ and $x = \frac{y - a_0}{a_1 + a_2 y}$, where $a_i = \hat{\alpha}_i$ in each respective model.

## 1.2 Non-Response Analysis

First introduced by Wooten in 2011, **non-response analysis** assumes $g(x, y) = 1$; hence, there is no initial constant $\alpha_0$, in $h(x, y)$. If $\{T_1(x, y), \ldots, T_m(x, y)\}$ are the remaining terms of interest, then the equation to be tested is

$$g(x, y) = 1$$

and

$$h(x, y|\theta) = \sum_{i=1}^{m} \alpha_i T_i(x, y),$$

and the system of equations becomes

$$\sum_{i=1}^{n} T_k(x_i, y_i) = \sum_{i=1}^{n} \sum_{j \neq k} \alpha_k T_k(x_i, y_i) T_j(x_i, y_i); k = 1, \ldots, m,$$

which no longer depends on the sample size but rather how each term relates in the system.

For example, if the terms of interest are $\{x, y, xy\}$, then the non-response model:

$$1 = \alpha_1 x + \alpha_2 y + \alpha_3 xy$$

which can be used to analyze the nature of the relationship that exist between the underlying variables in a co-dependent relationship; the system of equations take the form:

$$\sum_{i=1}^{n} x_i = \alpha_1 \sum_{i=1}^{n} x_i^2 + \alpha_2 \sum_{i=1}^{n} x_i y_i + \alpha_3 \sum_{i=1}^{n} x_i^2 y_i$$

$$\sum_{i=1}^{n} y_i = \alpha_1 \sum_{i=1}^{n} x_i y_i + \alpha_2 \sum_{i=1}^{n} y_i^2 + \alpha_3 \sum_{i=1}^{n} x_i y_i^2$$

$$\sum_{i=1}^{n} x_i y_i = \alpha_1 \sum_{i=1}^{n} x_i^2 y_i + \alpha_2 \sum_{i=1}^{n} x_i y_i^2 + \alpha_3 \sum_{i=1}^{n} x_i^2 y_i^2$$

Not only does this system of equations minimize the sum of square error in the system; it also follows directly from the given relationship illustrated in Figure 1; the blue represents the weight $\alpha_1$, red represents the weight $\alpha_2$ and black represents the weight $\alpha_3$. That is, the system of equations takes into account how each term affects the relationship with balancing weights.

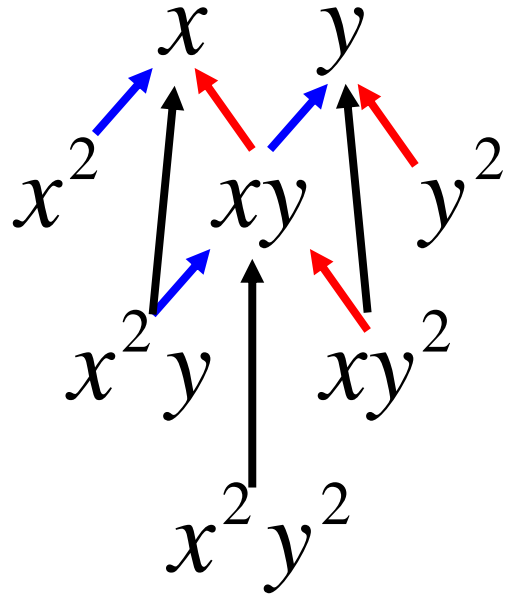

**Figure 1:** Diagram of the weighting system in the non-response model with terms of interest $\{x, y, xy\}$, $1 = \alpha_1 x + \alpha_2 y + \alpha_3 xy$. The weights balance the equations in terms of the directional arrows: $\alpha_1$ controls the contribution of $\sum_{i=1}^{n} x_i^2$ to $\sum_{i=1}^{n} x_i$, $\sum_{i=1}^{n} x_i y_i$ to $\sum_{i=1}^{n} y_i$, and $\sum_{i=1}^{n} x_i^2 y_i$ to $\sum_{i=1}^{n} x_i y_i$. Similarly, $\alpha_2$ controls the contribution of $\sum_{i=1}^{n} y_i^2$ to $\sum_{i=1}^{n} y_i$, $\sum_{i=1}^{n} x_i y_i$ to $\sum_{i=1}^{n} x_i$, and $\sum_{i=1}^{n} x_i y_i^2$ to $\sum_{i=1}^{n} x_i y_i$; and $\alpha_3$ controls the contribution of $\sum_{i=1}^{n} x_i^2 y_i$ to $\sum_{i=1}^{n} x_i$, $\sum_{i=1}^{n} x_i y_i^2$ to $\sum_{i=1}^{n} y_i$, and $\sum_{i=1}^{n} x_i^2 y_i^2$ to $\sum_{i=1}^{n} x_i y_i$.

Non-response analysis uses each term of interest as the random weighting structure (Jamison, Orey, Pruitt, 1965 & Etemadi, 2006 & Johnson, Kotzb 1990) on the relationship which can be used to detect non-functional relationships such as circles and ellipse. For example, if the terms of interest are $\{x, y, xy, x^2, y^2\}$, then the non-response model:

$$1 = \alpha_1 x + \alpha_2 y + \alpha_3 xy + \alpha_4 x^2 + \alpha_5 y^2$$

which can be used to analyze conical data. The resulting system of equations is given by

$$\begin{cases} \sum_{i=1}^{n} x_i = \alpha_1 \sum_{i=1}^{n} x_i^2 + \alpha_2 \sum_{i=1}^{n} x_i y_i + \alpha_3 \sum_{i=1}^{n} x_i^2 y_i + \alpha_4 \sum_{i=1}^{n} x_i^3 + \alpha_5 \sum_{i=1}^{n} x_i y_i^2 \\ \sum_{i=1}^{n} y_i = \alpha_1 \sum_{i=1}^{n} x_i y_i + \alpha_2 \sum_{i=1}^{n} y_i^2 + \alpha_3 \sum_{i=1}^{n} x_i y_i^2 + \alpha_4 \sum_{i=1}^{n} x_i^2 y_i + \alpha_5 \sum_{i=1}^{n} y_i^3 \\ \sum_{i=1}^{n} x_i y_i = \alpha_1 \sum_{i=1}^{n} x_i^2 y_i + \alpha_2 \sum_{i=1}^{n} x_i y_i^2 + \alpha_3 \sum_{i=1}^{n} x_i^2 y_i^2 + \alpha_4 \sum_{i=1}^{n} x_i^3 y_i + \alpha_5 \sum_{i=1}^{n} x_i y_i^3 \\ \sum_{i=1}^{n} x_i^2 = \alpha_1 \sum_{i=1}^{n} x_i^3 + \alpha_2 \sum_{i=1}^{n} x_i^2 y_i + \alpha_3 \sum_{i=1}^{n} x_i^3 y_i + \alpha_4 \sum_{i=1}^{n} x_i^4 + \alpha_5 \sum_{i=1}^{n} x_i^2 y_i^2 \\ \sum_{i=1}^{n} y_i^2 = \alpha_1 \sum_{i=1}^{n} x_i y_i^2 + \alpha_2 \sum_{i=1}^{n} y_i^3 + \alpha_3 \sum_{i=1}^{n} x_i y_i^3 + \alpha_4 \sum_{i=1}^{n} x_i^2 y_i^2 + \alpha_5 \sum_{i=1}^{n} y_i^4 \end{cases}$$

the internal weighting system is illustrated in Figure 2.

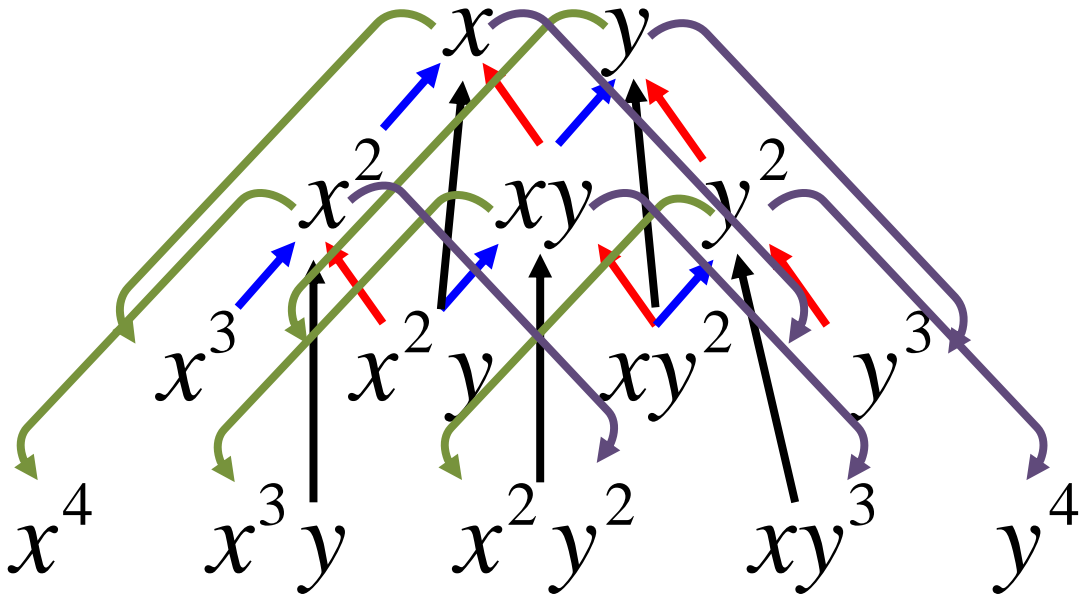

**Figure 2:** Diagram of the weighting system in the non-response model with terms of interest $\{x, y, xy, x^2, y^2\}$, $1 = \alpha_1 x + \alpha_2 y + \alpha_3 xy + \alpha_4 x^2 + \alpha_5 y^2$. The weights balance the equations in terms of the directional arrows: $\alpha_1$ controls the contribution of $\sum_{i=1}^{n} x_i T_j(x_i, y_i)$ to $\sum_{i=1}^{n} T_j(x_i, y_i)$. Similarly, $\alpha_2$ controls the contribution of $\sum_{i=1}^{n} y_i T_j(x_i, y_i)$ to $\sum_{i=1}^{n} T_j(x_i, y_i)$; $\alpha_3$ controls the contribution of $\sum_{i=1}^{n} x_i y_i T_j(x_i, y_i)$ to $\sum_{i=1}^{n} T_j(x_i, y_i)$; $\alpha_4$ controls the contribution of $\sum_{i=1}^{n} x_i^2 T_j(x_i, y_i)$ to $\sum_{i=1}^{n} T_j(x_i, y_i)$; and $\alpha_5$ controls the contribution of $\sum_{i=1}^{n} y_i^2 T_j(x_i, y_i)$ to $\sum_{i=1}^{n} T_j(x_i, y_i)$.

In the second order non-response model, both variables are tractable;

$$y = \frac{-(a_2 + a_3 x) \pm \sqrt{(a_2 + a_3 x)^2 - 4a_5(a_1 x + a_4 x^2 - 1)}}{2a_5}$$

and

$$x = \frac{-(a_1 + a_3 y) \pm \sqrt{(a_1 + a_3 x)^2 - 4a_4(a_2 y + a_5 y^2 - 1)}}{2a_4}$$

where $a_i = \hat{\alpha}_i$.

## 2.0  Origin of Implicit Regression

The idea behind **Implicit Regression** came about while considering the standard regression model where the subject response is not measured; that is, the subject response is a lurking variable. Let $z = h(x, y|\theta)$, where both $x$ and $y$ are considered explanatory variables, $z$ is the subject response, and $h(x, y)$ is defined in terms of the unknown coefficients, $\theta = \{\beta_0, \beta_1, \beta_2, \dots, \beta_m\}$. The subject response is assumed to be normally distributed; however, no assumptions are required for explanatory variables. In fact, in a good experimental design, varied outcomes in these variables are preferred; otherwise, they are relatively constant and are absorbed by the constant.

Wooten assumes that this response follows a normal probability distribution, $z \sim N(\mu, \sigma^2)$; however, the subject response is unknown or not measured. The construct behind this new methodology started with standard **Multi-linear Regression** (**MLR**) assuming that $h(x, y|\theta)$ is an additive model; that is, the relationship between the measured random variables is as follows:

$$z = \beta_0 + \beta_1 x + \beta_2 y + \beta_3 xy + \cdots \tag{1}$$

with random error in the observed data of the form

$$z_i = \beta_0 + \beta_1 x_i + \beta_2 y_i + \beta_3 x_i y_i + \cdots + \varepsilon_i, \qquad (2)$$

where $\beta_i$ are the parameters that weight the known explanatory variables, $\beta_0 \neq 0$, and $\varepsilon$ is the random error in the subject response $z$ is normally distributed with mean zero and constant variance, $\varepsilon \sim N(0, \sigma^2)$.

As $z$ in equation (2) is not measured, we cannot use standard regression analysis in the conventional way to detect this initial relationship. Then a combination of methods including Cramer's Rule (Cramer 1750, Muir 1960, Weinstein 2011), augmented matrices and aliases matrices came to mind. Solutions using Cramer's Rule and alias matrices as solutions to sub-model configurations is what will be referred to as non-response analysis and rotational analysis. Whereas ordinary least squares is used in standard regression analysis to minimize random error in one direction, rotational analysis minimizes random error in different directions; and non-response analysis minimizes random error in the system. Rotational analysis is a superior analysis that addressed implicitly defined relationship among co-dependent terms whereas standard regression analysis is for explicitly defined functions with an independent subject response.

### 2.1    Rotational Analysis as Aliases

**Rotational Analysis** is the testing of each terms effect on the remaining terms; that is, given equation (1), to measure the effects of the variable $x$ on the model, we consider the alias matrix, $A$, such that the expected value of the beta coefficients and the coefficient on the variable $x$ are related as follows

$$E\left(\begin{bmatrix}\beta_0\\\beta_2\\\vdots\end{bmatrix}\right) = \begin{bmatrix}\beta_0\\\beta_2\\\vdots\end{bmatrix} + A\beta_1,$$

where the alias is obtained by

$$A = (X_1'X_1)^{-1}X_1'X_2, \tag{3}$$

where $X_1 = \begin{bmatrix} 1 & y_1 & x_1y_1 & \cdots \\ \vdots & \vdots & \ddots & \vdots \\ 1 & y_n & x_ny_n & \cdots \end{bmatrix}$ and $X_2 = \begin{bmatrix} x_1 \\ \vdots \\ x_n \end{bmatrix}$. Testing the effects of x on the primary model

given in equation (1) is equivalent to testing the model

$$x = \alpha_0 + \alpha_1 y + \alpha_2 xy + \cdots, \tag{4}$$

where $A = \hat{\alpha}$, $E(\hat{\alpha}) = \alpha$ and $V(\hat{\alpha}) = \sigma^2(X_1'X_1)^{-1}$ and solution is the alias matrix given in equation (3).

For example, if the terms of interest are $\{x, y, xy, x^2, y^2\}$, then there are five rotations starting with the outlined model in equation (4) these are:

$$x = \alpha_0 + \alpha_1 y + \alpha_2 xy + \alpha_3 x^2 + \alpha_4 y^2$$

$$y = \alpha_0 + \alpha_1 x + \alpha_2 xy + \alpha_3 x^2 + \alpha_4 y^2$$

$$xy = \alpha_0 + \alpha_1 x + \alpha_2 y + \alpha_3 x^2 + \alpha_4 y^2,$$

$$x^2 = \alpha_0 + \alpha_1 x + \alpha_2 y + \alpha_3 xy + \alpha_4 y^2,$$

$$y^2 = \alpha_0 + \alpha_1 x + \alpha_2 y + \alpha_3 xy + \alpha_4 x^2,$$

which can be used to analyze the nature the relationship that exists between the underlying variables. In each case, we can test the value of each parameter estimates of the coefficients using the standard t-test and the overall model using the standard F-test. However, as we have violated the assumptions of independence, the sums of squares are not related as in standard regression; we will address this issue in the section on Model Evaluation. Note: in each of the above rotations, it is possible to use the quadratic equation to calculate estimates of both $x$ and $y$; namely, $\hat{x}$ and $\hat{y}$.

## 2.2. Non-Response Analysis as an Alias

**Non-response Analysis (NRA)** is the testing of model's constant on all the remaining terms; the alias matrix $A$, is such that the expected value of the beta coefficients and the constant coefficient are related as follows

$$E\left(\begin{bmatrix}\beta_1\\\beta_2\\\vdots\end{bmatrix}\right) = \begin{bmatrix}\beta_1\\\beta_2\\\vdots\end{bmatrix} + A\beta_0$$

measuring the bias the constant intercept has on all the remaining parameters using equation (3)

where $X_1 = \begin{bmatrix} x_1 & y_1 & x_1y_1 & \cdots \\ \vdots & \vdots & \vdots & \ddots \\ x_n & y_n & x_ny_n & \cdots \end{bmatrix}$ and $X_2 = \begin{bmatrix} 1 \\ \vdots \\ 1 \end{bmatrix}$; that is, this view shows the bias introduced by the constant and is equivalent to testing

$$1 = \alpha_1 x + \alpha_2 y + \alpha_3 xy + \alpha_4 x^2 + \alpha_5 y^2, \tag{5}$$

and $E(\hat{\alpha}) = \alpha$ and $V(\hat{\alpha}) = \sigma^2 (X_1'X_1)^{-1}$.

In the model outlined in equation (5), unity is treated as a random effect and not as a fixed effect.

Hence, implicit regression is the combination of rotational analysis and non-response analysis; and standard regression is a subset of rotational analysis and therefore a subset of implicit regression.

## 3.0 Standard Regression and Implicit Regression

Given a set of terms; for example, $\{x, y, xy, x^2, y^2\}$, the relationships that can be tested implicitly take the form $g(x,y) = h(x,y|\theta)$ whereas in standard regression, $g(x,y) = y$ and $h(x,y|\theta) = f(x|\theta)$ forcing a dependent/independent relationship with a 90° angle of separation.

### 3.1 Comparing Normal Equations in Standard Regression in MLR and NRA

Given the standard regression model with specified subject response, $y$, and independent explanatory variables, $x_1, x_2, \ldots, x_p$:

$$y_i = \beta_0 + \beta_1 x_{1i} + \cdots + \beta_p x_{pi} + \varepsilon_i, \tag{6}$$

with matrix representations of the independent measures

$$X = \begin{bmatrix} 1 & x_{11} & \cdots & x_{p1} \\ \vdots & \vdots & \ddots & \vdots \\ 1 & x_{1n} & \cdots & x_{pn} \end{bmatrix}$$

and the parameter matrix

$$\beta = \begin{bmatrix} \beta_0 \\ \vdots \\ \beta_{p-1} \\ \beta_p \end{bmatrix}$$

and response vector

$$Y = \begin{bmatrix} y_1 \\ \vdots \\ y_n \end{bmatrix}.$$

Then, we have the matrix equation

$$Y = X\beta + \varepsilon \tag{7}$$

we estimate the parameter using the normal equations given by

$$\hat{\beta} = (X'X)^{-1}X'Y$$

and the sums of squares are

$$\hat{Y}'\hat{Y} = Y'X(X'X)^{-1}X'Y$$

$$Y'Y - n\bar{y}^2 = \hat{Y}'\hat{Y} - n\bar{y}^2 + (Y - X\hat{\beta})'(Y - X\hat{\beta})$$

$$SST = Y'Y = \sum y^2, SSR = \hat{\beta}'X'Y, SSE = Y'Y - \hat{\beta}X'Y$$

with coefficient of determination

$$R^2 = \frac{\hat{\beta}'X'Y - n\bar{y}^2}{Y'Y - n\bar{y}^2}. \tag{8}$$

Here, the $R^2$ measures the percent variation in the response variable explained by the regression line above the mean and the explanatory variables; and is related to the number of the points used to develop the relationship in this standard form.

Now consider this model in the following manipulated form:

$$\beta_0 = y - \beta_1 x_1 - \cdots - \beta_p x_p$$

or, equivalently, assuming $\beta_0 \neq 0$,

$$1 = \alpha_0 y + \alpha_1 x_1 + \cdots + \alpha_p x_p, \tag{9}$$

or the non-response model, we have the augmented matrix $W = [Y|X]$; that is,

$$W = \begin{bmatrix} y_1 & x_{11} & \cdots & x_{p1} \\ \vdots & \vdots & \ddots & \vdots \\ y_n & x_{1n} & \cdots & x_{pn} \end{bmatrix}$$

and the parameter matrix is given by

$$\alpha = \begin{bmatrix} \alpha_0 \\ \alpha_1 \\ \vdots \\ \alpha_p \end{bmatrix}$$

and unity is the column of ones,

$$1 = 1_{1 \times n} = \begin{bmatrix} 1 \\ \vdots \\ 1 \end{bmatrix}.$$

Then, we have the non-response model is

$$W\alpha = 1 + \omega \tag{10}$$

with normal equations

$$\hat{\alpha} = (W'W)^{-1}W'1.$$

If $y$ is in fact the subject response, then the parameters outline in equations 6 and 9 hold the following relationships:

$$\beta_i = \frac{\alpha_i}{\alpha_0}; i = 1, \ldots, p$$

and

$$\beta_0 = -\frac{1}{\alpha_0}.$$

However, beyond standard regression, this idea can be extended to include higher order terms and interactions. Considered the augmented matrix $W = [X|Y|XY|X^2|Y^2 \cdots] = [Z]$, that is,

$$W = \begin{bmatrix} x_{11} & \cdots & x_{p1} & y_1 & x_{11}^2 & \cdots & x_{p1}^2 & x_{11}y_1 & \cdots & x_{p1}y_1 & y_1^2 \\ \vdots & \ddots & \vdots & \vdots & \vdots & \ddots & \vdots & \vdots & \ddots & \vdots & \vdots \\ x_{1n} & \cdots & x_{pn} & y_n & x_{1n}^2 & \cdots & x_{pn}^2 & x_{1n}y_n & \cdots & x_{pn}y_n & y_n^2 \end{bmatrix} = \begin{bmatrix} z_{11} & \cdots & z_{k1} \\ \vdots & \ddots & \vdots \\ z_{1n} & \cdots & z_{kn} \end{bmatrix}$$

and the parameter matrix

$$\alpha = \begin{bmatrix} \alpha_1 \\ \vdots \\ \alpha_k \end{bmatrix}$$

where $k = 3p + 2$ for a full second order relation.

In this relationship, we have

$$W_{n \times k} \alpha_{k \times 1} = 1_{n \times 1} \quad (11)$$

$$\hat{\alpha} = (W'W)^{-1} W'1$$

with sums of squares given by

$$\hat{1}'\hat{1} = 1'W(W'W)^{-1}W'1$$

$$1'1 = \hat{1}'\hat{1} + (1 - W\hat{\alpha})'(1 - W\hat{\alpha})$$

$$SST = 1'1 = n, SSR = \hat{\alpha}'W'1, SSE = n - \hat{\alpha}'W'1$$

and the coefficient of determination becomes

$$R^2 = \frac{\hat{\alpha}'W'1}{n} = \frac{1'W(W'W)^{-1}W'1}{n}. \quad (12)$$

Here, the $R^2$ in equation (12) is not comparable to equation (8) as the relationship between the sums of squares no longer hold; that is, the degree of separation is not 90° and therefore $SST \neq SSR + SSE$. A measure for the degree of separation is given in the section on Model Evaluation.

### 3.2 Usefulness of New Theory in Univariate and Bivariate Analysis

Implicit regression is useful in multiple linear regression to analysis the relations that exist between the variables; however, here the usefulness is illustrated for univariate and bivariate analysis.

### 3.2.1 Univariate Analysis

Considered the univariate model $1 = \alpha y$ and the matrix information for the subject response

$$Y = \begin{bmatrix} y_1 \\ \vdots \\ y_n \end{bmatrix}$$

and the parameter matrix

$$\alpha = [\alpha_0]$$

and unity is

$$1 = \begin{bmatrix} 1 \\ \vdots \\ 1 \end{bmatrix}.$$

Then, we have

$$Y\alpha = 1 + \omega, \tag{13}$$

where $\omega \sim N(0, \sigma^2)$, and normal equations given by

$$\hat{\alpha} = (Y'Y)^{-1}Y'1 = \frac{\sum y_i}{\sum y_i^2}$$

and

$$R^2 = \frac{1'Y(Y'Y)^{-1}Y'1}{n} = \frac{(\sum y_i)^2}{n \sum y_i^2}. \tag{14}$$

This is the same as in standard regression for univariate; without correction for the mean, $R^2$ measures the percent variation in the response variable (sums of squares) explained by the sample mean above the zero measure and is given by

$$R^2 = \frac{\hat{\beta}'X'Y}{Y'Y} = \frac{Y'X(X'X)^{-1}X'Y}{Y'Y} = \frac{(\sum y)\frac{1}{n}(\sum y)}{\sum y^2} = \frac{(\sum y_i)^2}{n \sum y_i^2}.$$

This measure $R^2$ in equation (14) is an indication of the **constant nature of the variable** itself.

Consider the sums of square error for the variable $y$,

$$\sum (y - \bar{y})^2 = \sum y^2 - n\bar{y}^2$$

then we have

$$\sum y^2 = \sum (y - \bar{y})^2 + n\bar{y}^2.$$

Hence, internal to the variable, the coefficient of determination is the percent of total sums of squares explained by the mean.

$$R^2 = \frac{n\bar{y}^2}{\sum y^2} = \frac{(\sum y_i)^2}{n \sum y_i^2}.$$

As $R^2 \to 1$, $V(y) \to 0$ and therefore $y$ is approximately a constant.

As $R^2 \to 0$, $V(y) \to \infty$ and therefore $y$ is extremely variant.

*Normally Distributed Random Variables*

If $y \sim N(\mu, \sigma^2)$, then $E(\bar{y}) = \mu \neq 0$ and $E(\sum y^2) = n\mu^2 + \sigma^2$

$$R^2 = \frac{n\mu^2}{n\mu^2 + \sigma^2}.$$

Therefore, assuming constant mean, $R^2 \to 1$ if either $\sigma^2 \to 0$ or $n \to \infty$; and $R^2 \to 0$ only if $n\mu^2 + \sigma^2 \to \infty$ and $n\mu^2 \not\to \infty$, which implies $R^2 \to 0$ if and only if $\sigma^2 \to \infty$.

If $y \sim U(a, b)$, then $E(\bar{y}) = \frac{a+b}{2}$ and $E(\sum y^2) = n\frac{b^2+ab+a^2}{3}$

$$R^2 = \frac{n\left(\frac{a+b}{2}\right)^2}{n\frac{b^2+ab+a^2}{3}} = \frac{3(a+b)^2}{4(b^2+ab+a^2)}.$$

Therefore, $R^2 \to 1$ only if $a \to b$ independent of sample size; and assuming $a$ is a fixed point,

$R^2 \to 0$ only if $b = -a$; however, if $a + b \to \infty$, this implies $R^2 \to \frac{3}{4}$.

Note: This value is related to the coefficient of variation, $CV = \frac{\sigma}{\mu}$; as $CV \to 0, \sigma \to 0$ and $R^2 \to 1$.

However, for uniformly distributed data, $CV = \frac{\sqrt{3}}{3}\frac{(b-a)}{(b+a)} \to \frac{\sqrt{3}}{3}$ as $b \to \infty$.

### 3.2.2 Usefulness in Bivariate Analysis

Consider **Simple Linear Regression** (**SLR**) with data matrices:

$$X = \begin{bmatrix} 1 & x_1 \\ \vdots & \vdots \\ 1 & x_n \end{bmatrix}$$

and response array

$$Y = \begin{bmatrix} y_1 \\ \vdots \\ y_n \end{bmatrix}.$$

and the parameter matrix

$$\beta = \begin{bmatrix} \beta_0 \\ \beta_1 \end{bmatrix}$$

Then we have

$$\Delta = n\sum x^2 - \left(\sum x\right)^2$$

$$\hat{\beta} = \frac{1}{\Delta}\begin{bmatrix} \sum x^2 & -\sum x \\ -\sum x & n \end{bmatrix}\begin{bmatrix} \sum y \\ \sum xy \end{bmatrix}$$

$$\hat{\beta} = \frac{1}{\Delta}\begin{bmatrix} \sum y \sum x^2 - \sum x \sum xy \\ n \sum xy - \sum x \sum y \end{bmatrix}$$

which given

$$y = \beta_0 + \beta_1 x$$

and

$$y_i = \beta_0 + \beta_1 x + \varepsilon_i. \tag{16}$$

minimizes the sum of square errors, that is, the vertical distance between the point and the line, $\varepsilon_i^2 = (y_i - \beta_0 - \beta_1 x)^2$; thus, we have

$$b_1 = \hat{\beta}_1 = \frac{n\sum xy - \sum x \sum y}{n\sum x^2 - (\sum x)^2}$$

and

$$b_0 = \hat{\beta}_0 = \bar{y} - \hat{\beta}_1 \bar{x}.$$

which takes into account the effects of the sample size, $n$, on $y$ and the effects of the variable $x$ on $y$.

In **non-response analysis**, we considered the augmented matrix $W = [X|Y]$, that is,

$$W = \begin{bmatrix} x_1 & y_1 \\ \vdots & \vdots \\ y_n & y_n \end{bmatrix}$$

and the parameter matrix

$$\alpha = \begin{bmatrix} \alpha_0 \\ \alpha_1 \end{bmatrix}$$

and unity is

$$1 = \begin{bmatrix} 1 \\ \vdots \\ 1 \end{bmatrix}.$$

Then we have

$$W\alpha = 1$$

with normal equations

$$\hat{\alpha} = (W'W)^{-1}W'1$$

which is equivalent to computing the determinant, $\Delta$, and using inverse matrices to estimate the parameter:

$$\Delta = \sum x^2 \sum y^2 - \left(\sum xy\right)^2$$

$$\hat{\alpha} = \frac{1}{\Delta} \begin{bmatrix} \sum y^2 & -\sum xy \\ -\sum xy & \sum x^2 \end{bmatrix} \begin{bmatrix} \sum x \\ \sum y \end{bmatrix}$$

or

$$\hat{\alpha} = \frac{1}{\Delta} \begin{bmatrix} \sum y^2 \sum x - \sum xy \sum y \\ \sum x^2 \sum y - \sum xy \sum x \end{bmatrix}$$

which given

$$1 = \alpha_1 x + \alpha_2 y$$

and

$$1 = \alpha_1 x_i + \alpha_2 y_i + \omega_i$$

minimizes the sum of square errors, that is, the distance between the point and the line, $\omega_i^2 = (1 - \alpha_1 x_i - \alpha_2 y_i)^2$; thus, we have

$$a_1 = \hat{\alpha}_1 = \frac{\sum y^2 \sum x - \sum xy \sum y}{\sum x^2 \sum y^2 - (\sum xy)^2}$$

and

$$a_2 = \hat{\alpha}_2 = \frac{\sum x^2 \sum y - \sum xy \sum x}{\sum x^2 \sum y^2 - (\sum xy)^2}$$

which does not account for the effects of the sample size, $n$, but rather accounts for the effects between the variable $x$ on $y$ with coefficient of determination

$$R^2 = \frac{\hat{\alpha}' W' 1}{n} = \frac{1' W (W'W)^{-1} W' 1}{n}.$$

Estimating the parameter using this procedure is similar to minimizing the distance between the point and line; addressing random error in both variables simultaneously.

### 4.0 Comparing OLS for Univariate and Bivariate Analysis

Both standard regression and implicit regression, ordinary least squares is used to determine the parameter estimates.

## 4.1 OLS in Univariate Analysis

In standard regression (Bulmer, 2003), we have the assumption that $y \sim N(\beta = \mu, \sigma^2)$ and

$$y = \beta$$

with observed data that is related to the random error as follows:

$$y_i = \beta + \varepsilon_i,$$

where $E(\varepsilon) = 0$ and $V(\varepsilon) = \sigma^2$; that is, $\varepsilon \sim N(0, \sigma^2)$.

Therefore, to minimize the sum of square errors, we have

$$Q = \sum (\varepsilon)^2 = \sum (y_i - \beta)^2$$

$$\frac{\partial Q}{\partial \beta} = \sum 2(y_i - \beta) = 0$$

$$\hat{\beta} = \frac{\sum y_i}{n} = \hat{\mu}_{y\cdot},$$

which relates the total sum of the data to the sample size.

In non-response analysis, we consider

$$\alpha y = 1$$

with observed data is related to the random error as follows:

$$\alpha y_i = 1 + \omega_i$$

where $E(\omega) = 0$ and $V(\omega) = V(\alpha y - 1) = V\left(\frac{y-\mu}{\mu}\right) = \frac{\sigma^2}{\mu^2}$; that is, $\omega \sim N\left(0, \frac{\sigma^2}{\mu^2}\right)$.

Therefore, to minimize the sum of square error as measured using the percent difference, we have

$$Q = \sum (\omega)^2 = \sum (\alpha y - 1)^2$$

$$\frac{\partial Q}{\partial \alpha} = \sum 2(\alpha y - 1)y = 0$$

$$\hat{\alpha} = \frac{\sum y}{\sum y^2}.$$

$$\hat{\mu}_y = \frac{1}{\hat{\alpha}} = \frac{\sum y^2}{\sum y},$$

which relates the total sums of squares to the sum of the data; the self-weighting mean that treats the measure *y* as both a variable and constant weights.

Note: this is a biased estimate in that, $E(\hat{\mu}_y) = \mu_y + \frac{\sigma_y^2}{\mu_y}$ which converges to $\mu_y$ as $\mu_y \to \infty$ or as $\sigma_y^2 \to 0$. That is, if the mean is significantly greater than 0 or the measure is rather constant with small variance, the this estimate of the mean has small bias. The variance is the coefficient of variation (the ratio of the standard deviation and the mean).

### 4.2 OLS in Bivariate Analysis and Euclidian Distance

Given the non-response model with two linearly related measures,

$$\alpha_1 x + \alpha_2 y = 1$$

and the distance between a point and this line is given by

$$D_i = \frac{|\alpha_1 x_i + \alpha_2 y_i - 1|}{\sqrt{\alpha_1^2 + \alpha_2^2}} = |\gamma_1 x_i + \gamma_2 y_i - \gamma|,$$

where $\gamma_1 = \frac{\alpha_1}{\sqrt{\alpha_1^2+\alpha_2^2}}$, $\gamma_2 = \frac{\alpha_2}{\sqrt{\alpha_1^2+\alpha_2^2}}$. and $\gamma = \frac{1}{\sqrt{\alpha_1^2+\alpha_2^2}}$; and to minimize the sum of square distances,

that is, the perpendicular distance between the point and the line, we have

$$Q = \sum D_i^2 = \sum (\gamma_1 x_i + \gamma_2 y_i - \gamma)^2,$$

and we obtain three equations by taking the partial derivatives with respect to each parameter:

$$\frac{\partial Q}{\partial \gamma_1} = \sum 2(\gamma_1 x_i + \gamma_2 y_i - \gamma)(x_i) = 0$$

$$\frac{\partial Q}{\partial \gamma_2} = \sum 2(\gamma_1 x_i + \gamma_2 y_i - \gamma)(y_i) = 0$$

$$\frac{\partial Q}{\partial \gamma} = \sum 2(\gamma_1 x_i + \gamma_2 y_i - \gamma)(1) = 0$$

which simplifies to

$$\gamma_1 \sum x_i^2 + \gamma_2 \sum x_i y_i = \gamma \sum x_i$$

$$\gamma_1 \sum x_i y_i + \gamma_2 \sum y_i^2 = \gamma \sum y_i$$

$$\gamma_1 \sum x_i + \gamma_2 \sum y_i = n\gamma$$

which is equivalent to solving

$$\alpha_1 \sum x_i^2 + \alpha_0 \sum x_i y_i = \sum x_i$$

$$\alpha_1 \sum x_i y_i + \alpha_0 \sum y_i^2 = \sum y_i.$$

same as above.

In simple linear regression the relationship spans the sample

$$\sum_{i=1}^{n}\left(-\frac{b_1}{b_0}x_i + \frac{1}{b_0}y_i\right) = n$$

whereas in non-response analysis,

$$\sum_{i=1}^{n}(a_1 x_i + a_2 y_i) \leq n,$$

and

$$R^2 = \frac{\sum_{i=1}^{n}(a_1 x_i + a_2 y_i)}{n}.$$

If there is no error in the system, then $\sum_{i=1}^{n}(a_1 x_i + a_2 y_i) = n$.

If the relationship is truly linear with small system error, then the non-response model is in line with the two rotations, Figure 3a; however, the greater the system error, the non-response model deviates from the rotations, Figure 3b; and when the system is non-linear, then the resulting lines have a pin-wheel effect, Figure 3c.

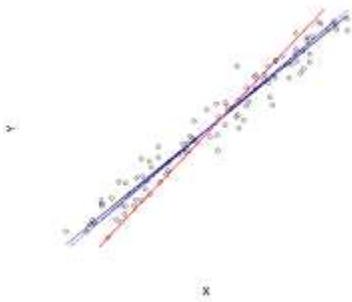 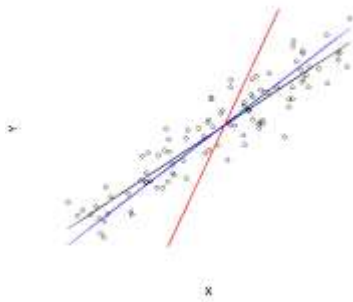 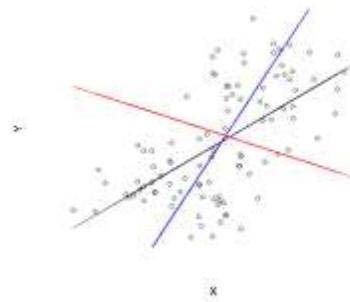

**Fig.3a**  **Fig.3b**  **Fig.3c**
**Figure 3:** Scatter plot of (a) linearly related variables with small variance, (b) a linearly related variables with double the variance, and (c) non-linear variables.

The evaluation of the model depends on the tractability of the solutions. Regardless of whether we can solve the implicit equation for $y$ (or both $x$ and $y$), we can measure the coefficient of determination; however, in non-response analysis, it measures the percent of the sample in the system explained by the extraneous variables. To evaluate the percent variance in $y$, we would first solve for $y$ in the relationship and use this function to estimate the subject response, $\hat{y}$, and then consider the relationship between sum of square total, $SST = \sum_{i=1}^{n}(y_i - \bar{y})^2$, the sum of square error between the model and the mean, $SSM = \sum_{i=1}^{n}(\hat{y}_i - \bar{y})^2$, and sum of the squares error unexplained between the model and the data, $SSE = \sum_{i=1}^{n}(\hat{y}_i - y_i)^2$. In standard regression, $SST = SSM + SSE$, and the angle $\theta_T$ between the $M$ and $E$ in the vector space is 90°, illustrated in Figure 4.

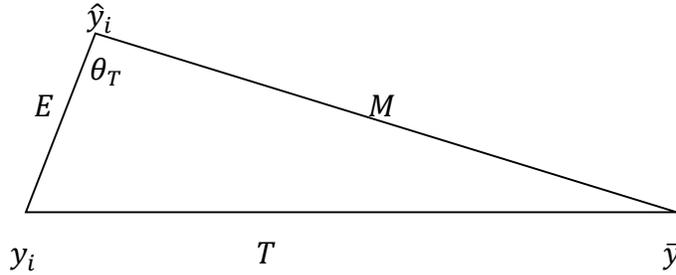

**Figure 4:** Graphical representation of the observed data, $y_i$, the grand mean, $\bar{y}$, and the estimated value of subject response, $\hat{y}_i$. The total error, $T_i = y_i - \bar{y}$ is shown with the degree of separation $\theta_T$; explained error, $M_i = \hat{y}_i - \bar{y}$, and unexplained error, $E_i = y_i - \hat{y}_i$.

However, as the assumption of independents is not satisfied and the degree of separation, $\theta_T$, is not guaranteed to be 90°; hence, we invoke the law of cosines to measure $\theta_T$. In general, we have

$$SST = SSM + SSE - 2\sqrt{SSM \times SSE}\cos(\theta_T),$$

which can be manipulated to be

$$\theta_T = \arccos\left(\frac{SSM + SSE - SST}{2\sqrt{SSM \times SSE}}\right),$$

where the closer $\theta_T$ is to $90°$, the better the model teases out the dependent/independent relationship.

To evaluate the model in terms of both $x$ and $y$, we would first solve for $y$ in the relationship and use this function to estimate the subject response, $\hat{y}$, and then solve for $x$ in the relationship and use this function to estimate the subject response, $\hat{x}$; then consider the relationship between the sum of square errors in bivariate measures,

$$SST = \sum_{i=1}^{n}(y_i - \bar{y})^2 + \sum_{i=1}^{n}(x_i - \bar{x})^2,$$

the sum of square error that exist between the two resulting models and the means,

$$SSM = \sum_{i=1}^{n}(\hat{y}_i - \bar{y})^2 + \sum_{i=1}^{n}(\hat{x}_i - \bar{x})^2,$$

and sum of the squares error that is unexplained between the model and the data,

$$SSE = \sum_{i=1}^{n}(\hat{y}_i - y_i)^2 + \sum_{i=1}^{n}(\hat{x}_i - x_i)^2.$$

As illustrated below, implicit regression allows us to extend the vector space to be bivariate and measure $\theta_T$ and all other **degrees of separation**: $\theta_M$ and $\theta_E$, including the height, $h$, or the extent to which the point estimates are removed from the data and the means, Figure 5.

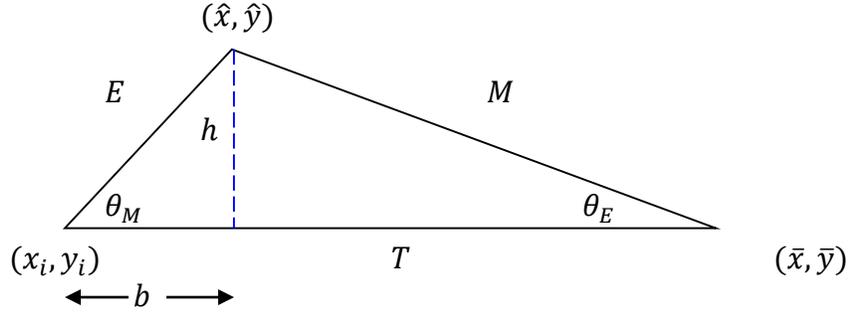

**Figure 5:** Graphical representation of the observed data, $(x_i, y_i)$, the grand means, $(\bar{x}, \bar{y})$, and the estimated value of variables of interest, $(\hat{x}_i, \hat{y}_i)$. The total error is the Euclidian distance, $T_i^2 = (y_i - \bar{y})^2 + (x_i - \bar{x})^2$ is shown with the height $h$; with explained error, $M_i^2 = (\hat{y}_i - \bar{y})^2 + (\hat{x}_i - \bar{x})^2$ and unexplained error, $E_i^2 = (y_i - \hat{y}_i)^2 + (x_i - \hat{x}_i)^2$.

The measured angles, $\theta_M$, where $\theta_M$ is the angle between $SST$ and $SSE$ given as

$$SSM = SST + SSE - 2\sqrt{SST \times SSE}\cos(\theta_M),$$

and

$$\theta_M = \arccos\left(\frac{SST + SSE - SSM}{2\sqrt{SST \times SSE}}\right),$$

and the **height** or extent to which the estimates are removed from the data and the mean is given by

$$h = \hat{E}\sin(\theta_M)$$

where $\hat{E} = \sqrt{\frac{SSE}{n}}$. A good model should have an angle close to 90° with height $h$, close to the ratio $\frac{ME}{T}$; that is, in a right triangle, $Ratio = \frac{hT}{ME} = 1$ which can be estimated using

$$Ratio = \sqrt{\frac{SST}{SSM}}\sin\theta_M.$$

The closer this ratio is to one and the closer the degree of separation, $\theta_T$, is to 90°, the better the developed model teases out the true relationship among the measured variables.

## 6. Conclusion

Ordinary least squares can be used to minimize the error in a specific direction, in a given term or in the overall system. Rotating the terms into the subject response position, the co-dependent relationship can be determine including the non-response model which places unity (a vector column of ones) as the subject response. Comparable to standard regression when terms placed in the subject response position are limited to one variable as a function of strictly independent variables, implicit regression houses standard independent/dependent relationships in addition to co-dependent relationships. Able to detect circles and ellipse and other tractable relationship, the measure defined as **degree of separation** formed by the implicitly defined relationship assesses how well the model outlines the independent/dependent or co-dependent relationship.